\documentclass[conference, a4paper]{IEEEtran}
\IEEEoverridecommandlockouts

\usepackage[sort]{cite}

\ifCLASSINFOpdf
  \usepackage[pdftex]{graphicx}
   \graphicspath{{./figures/}}
 
 \else

  \usepackage[dvips]{graphicx}
\fi

\usepackage{array}
\usepackage{booktabs}
\usepackage{amssymb}
\usepackage{amsmath}
\usepackage{balance}
\DeclareMathOperator*{\argmin}{argmin}

\usepackage{tikz}
\usetikzlibrary{positioning}
\usepackage[pscoord]{eso-pic}
\newcommand{\placetextbox}[3]{
\setbox0=\hbox{#3}
\AddToShipoutPictureFG*{
\put(\LenToUnit{#1\paperwidth},\LenToUnit{#2\paperheight}){\vtop{{\null}\makebox[0pt][c]{#3}}}%
}%
}%

\begin{document}

\title{Predicting Preferred Dialogue-to-Background Loudness Difference in Dialogue-Separated Audio}

\author{
    \IEEEauthorblockN{Luca Resti\IEEEauthorrefmark{1}, Martin Strauss\IEEEauthorrefmark{2}, Matteo Torcoli\IEEEauthorrefmark{1}, Emanu\"{e}l Habets\IEEEauthorrefmark{2}, Bernd Edler\IEEEauthorrefmark{2}}
    \IEEEauthorblockA{\IEEEauthorrefmark{1}Fraunhofer Institute for Integrated Circuits (IIS), Erlangen, Germany
    }
    \IEEEauthorblockA{\IEEEauthorrefmark{2}International Audio Laboratories Erlangen, Germany \thanks{$^\dag$A joint institution of the Friedrich-Alexander-Universit{\"a}t Erlangen-N{\"u}rnberg (FAU) and Fraunhofer Institute for Integrated Circuits (IIS).}
    }
}

\maketitle

\begin{tikzpicture}[overlay, remember picture]
    \path (current page.north) node (anchor) {};
    \node [below=of anchor] {%
    2023 15th International Conference on Quality of Multimedia Experience (QoMEX)};
\end{tikzpicture}

\vspace{-5pt}
\begin{abstract}
Dialogue Enhancement (DE) enables the rebalancing of dialogue and background sounds to fit personal preferences and needs in the context of broadcast audio.  When individual audio stems are unavailable from production, Dialogue Separation (DS) can be applied to the final audio mixture to obtain estimates of these stems. 
This work focuses on Preferred Loudness Differences (PLDs) between dialogue and background sounds. While previous studies determined the PLD through a listening test employing original stems from production, stems estimated by DS are used in the present study. In addition, a larger variety of signal classes is considered. PLDs vary substantially across individuals (average interquartile range: $\mathbf{5.7}\mskip3mu$LU). 
Despite this variability, PLDs are found to be highly dependent on the signal type under consideration, and it is shown that median PLDs can be predicted using objective intelligibility metrics. Two existing baseline prediction methods - intended for use with original stems - displayed a Mean Absolute Error (MAE) of $\mathbf{7.5}\mskip3mu$LU and $\mathbf{5}\mskip3mu$LU, respectively. A modified baseline (MAE: $\mathbf{3.2}\mskip3mu$LU) and an alternative approach (MAE: $\mathbf{2.5}\mskip3mu$LU) are proposed. Results support the viability of processing final broadcast mixtures with DS and offering an alternative remixing that accounts for median PLDs.
\end{abstract}

\placetextbox{0.5}{0.08}
{\fbox{\parbox{\dimexpr\textwidth-2\fboxsep-2\fboxrule\relax}{\footnotesize \centering Accepted paper. \copyright  2023 IEEE. Personal use of this material is permitted. Permission from IEEE must be obtained for all other uses, in any current or future media, including reprinting/republishing this material for advertising or promotional purposes, creating new collective works, for resale or redistribution to servers or lists, or reuse of any copyrighted component of this work in other works.}}}

\IEEEpeerreviewmaketitle

\vspace{-5pt}
\section{Introduction}
Finding the appropriate balance between speech in the Foreground (FG), also referred to as dialogue, and sounds in the Background (BG) in TV productions is often challenging. Optimal dialogue-to-background Loudness Differences (LDs) have been shown to be highly subjective and situation-specific \cite{geary2020loudness}, with high relative BG levels increasing Listening Effort (LE) for audiences, even when speech remains fully intelligible \cite{klink2012measuring}. Conversely, BG sounds should not be overly attenuated, as they lend narrative cues to a given scene. In traditional broadcast audio production, audio engineers manually balance audio stems, producing a fixed mixture of FG and BG sounds. Instead of static mixtures, Object-Based Audio (OBA) systems such as \mbox{MPEG-H} Audio \cite{mpeghstandard} are capable of representing individual audio sources, making their relative loudness levels adjustable during playback on end-user devices.
Thereby, OBA also enables Dialogue Enhancement (DE), i.e., allowing the audience to adjust relative dialogue levels to improve the overall quality of experience \cite{ast} and reduce LE \cite{9900884}.
However, it is not uncommon for broadcasters to have access only to the final stereo or 5.1 soundtracks. This makes it impossible to perform DE using the original production stems. In this case, individual FG and BG signals must be estimated via Dialogue Separation (DS). There are a few methods specifically targeting this application. While early works primarily implement non-Deep-Neural-Network (DNN) based approaches \cite{uhle2008speech, geiger_2015, paulus2019source}, DNNs have now emerged as the state-of-the-art for DS \cite{strauss21_interspeech, Westhausen2021, cocktailfork2022, paulus2022sampling}.

Our research questions are 1) what are the audience-preferred LDs between the estimated FG and BG, and 2) whether and how LDs can be predicted. In this work, LDs are expressed in Loudness Units (LU), with a relative level increase of $1\mskip3mu$dB corresponding to a $1\mskip3mu$LU increase. In \cite{tang2018automatic}, it was proposed to use the Binaural Distortion-Based Weighted Glimpse Proportion (BiDWGP) \cite{tang2015glimpse} Objective Intelligibility Metric (OIM) to predict experimentally-determined LDs. The BiDWGP metric was selected as it outperforms competing metrics in both anechoic \cite{tang2016metricanechoic} and reverberant \cite{tang2016evaluatingreverberant} conditions. LDs that yield a BiDWGP score of $0.5$ with an offset of $5.5\mskip3mu$dB were used in \cite{tang2018automatic}, with \cite{torcoli2019preferred} later showing that these parameters lead to an underestimation of the preferred LDs. An offset of $17.7\mskip3mu$LU resulted in LDs that better matched listener preferences.

Our work follows up on these results by investigating Preferred Loudness Differences (PLDs) considering potentially distorted dialogue-separated signals, which has not been previously considered. In addition, a larger variety of items is considered in this work, distinguishing between speech recorded in a studio and in less ideal acoustic conditions.

\section{DNN-Based Dialogue Separation}
\label{section:DNNDS}

The input mixture \textbf{y} $\in \mathbb{R}^{N \times 2}$ is defined as consisting of a FG  \textbf{d} $\in \mathbb{R}^{N \times 2}$ and a BG component \textbf{b} $\in \mathbb{R}^{N \times 2}$, i.e., $\mathbf{y} = \mathbf{d} + \mathbf{b}$ where $N$ denotes the length of the components.
DS aims to estimate the underlying FG and BG components from the mixture. To obtain the dialogue-enhanced mixture $\hat{\textbf{y}}$ the estimated components $\hat{\textbf{d}}$ and $\hat{\textbf{b}}$ are mixed back together with an attenuation factor $\beta$ on the estimated BG, i.e.,  $\hat{\mathbf{y}} = \hat{\mathbf{d}} + \beta \, \hat{\mathbf{b}}$.

The DS model used in this work has the same architecture as the 48\,kHz CNN described in \cite{paulus2022sampling}. The network operates in the complex STFT domain, is fully convolutional and has approximately 360\,000 trainable parameters.

\vspace{-2pt}
\section{Listening Test}
A listening test was conducted to determine what constitutes a PLD. In \cite{torcoli2019preferred}, the authors focused exclusively on two classes of broadcast items, namely Commentary over Music (CoM) and Commentary over Ambience (CoA), where \textit{commentary} refers to speech recorded in a studio by professionals who speak clearly. A typical example of commentary is the narration in a documentary. In \cite{tang2018automatic}, speech signals were not classified based on their recording conditions. In contrast, the present work also considers on-set Dialogue over Music (DoM) and on-set Dialogue over Ambience (DoA). \textit{On-set Dialogue} refers to speech recorded in less-than-ideal acoustic conditions, e.g., directly on the set of a narrative movie and with less clear enunciation by the speakers. Taken together, these item classes represent a large portion of audio found in TV productions. Differences in PLDs between on-set dialogue and commentary are examined in Section~\ref{section:testResults}.

\begin{figure}[t!]
\includegraphics[width=\columnwidth]{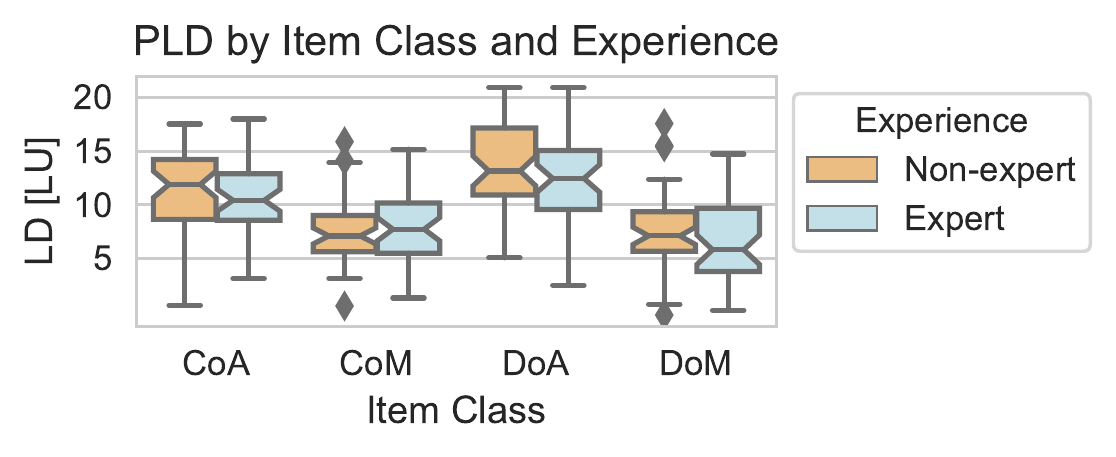}
\vspace{-20pt}
\caption{Boxplot of Preferred Loudness Differences (PLDs) of the four classes of test items for expert and non-expert subjects. A higher Loudness Difference (LD) indicates greater Background (BG)  attenuation. (Best viewed in color.)}
\vspace{-13pt}
\label{fig:pldByBgAndExp}
\end{figure}

\subsection{Participants and Test Setup}

The listening test involved $20$ German-speaking subjects aged $23$ to $48$ (median: $27$), with 18 subjects being under 31 years of age. Ten were classed as experts and ten as non-experts. Experts had prior experience with listening tests, while all non-experts had never taken part in a listening test before. A total of 20 excerpts ($9$ to $14$~s long) from real German broadcast TV content were used. The test items featured an equal number of male and female speakers and a sampling rate of $48\mskip3mu$kHz.
The test items were processed by the DNN-based DS system described in Section~\ref{section:DNNDS}. The estimated signal components were then mixed into a single stereophonic signal to produce eight conditions per item, which were presented in a randomized order. The estimated FG was loudness normalized to $-23$ LU full scale \cite{ebur128} - while the estimated BG was progressively attenuated in steps of $3\mskip3mu$LU, starting at a relative difference of $0\mskip3mu$LU, up to $21\mskip3mu$LU.

Subjects were shown a Multiple Stimuli with Hidden Reference and Anchor (MUSHRA) \cite{mushra1534} style interface through the listening test suite webMUSHRA \cite{schoeffler2018webmushra}. For every test item, subjects could listen to all eight conditions and rate them on a $0$ to $100$ scale. Participants were instructed to rate the mixes of FG and BG sounds based on their overall preference as if they were watching a TV program at home. The test was preceded by a trial phase to familiarize subjects with the test protocol. Listening preferences were studied in terms of the PLDs, as described in \cite{torcoli2019preferred}. For each test item and subject, the PLD is defined as the LD corresponding to the highest-rated condition. If more than one condition receives the highest rating, the mean of the LDs is considered instead. This approach bypasses the impact of each participant's different uses of the rating scale. 

Finally, measuring LDs directly on the estimated stems may not reflect the actual LD available to test subjects. Similarly to \cite{ast}, we estimated actual LDs with the help of the BSS Eval Toolbox \cite{vincent2006performance}. The mix modified by DS is decomposed into FG, BG and artifact components by the BSS Eval projection. The item-dependent LD is then calculated as the difference between the integrated loudness of the projected FG and of the projected BG. Similarly to \cite{netflixDGating}, the integrated loudness is computed by first applying dialogue-gating (segments where dialogue is not active are removed) and then applying the measurement in BS.1770-1 \cite{bs17701}. Dialogue gating is done because LDs only yield meaningful results when both FG and BG sounds are simultaneously active.

\begin{table}[t!]
\centering
\renewcommand{\arraystretch}{1.2}
\caption{ANOVA of the preferred loudness differences. Factors ordered by decreasing effect size $\eta^2$ (expressed as a percentage of total variability). Statistically significant factors ($\alpha = 0.05$) are marked in bold.}

\begin{tabular}{lcc}
\toprule
\textbf{Variable}    & \textbf{p} & \boldmath$\eta^2 \hspace{0.5em} (\%)$	 \\ \midrule
\textbf{Participant}                   & \textless$0.05$\hphantom{\textless}               & 32.31\\
\textbf{Item}                          & \textless$0.05$\hphantom{\textless}               & 27.18\\
\textbf{Foreground-Background Type}    & \textless$0.05$\hphantom{\textless}               & 24.15\\
\textbf{Participant x Foreground-Background Type} & \textless$0.05$\hphantom{\textless}               & \hphantom{0}4.92\\
\textbf{Experience}                    & \textless$0.05$\hphantom{\textless}               & \hphantom{0}1.69\\
Item x Experience             & \hphantom{\textless}$0.38$\hphantom{\textless}               & \hphantom{0}0.67\\
Residual                      &                     & \hphantom{0}9.08\\
\bottomrule
\end{tabular}
\vspace{-3pt}
\label{table:anovaTable}
\end{table}

\begin{figure*}[t!]
\centering
\includegraphics[width=\textwidth, scale=1]{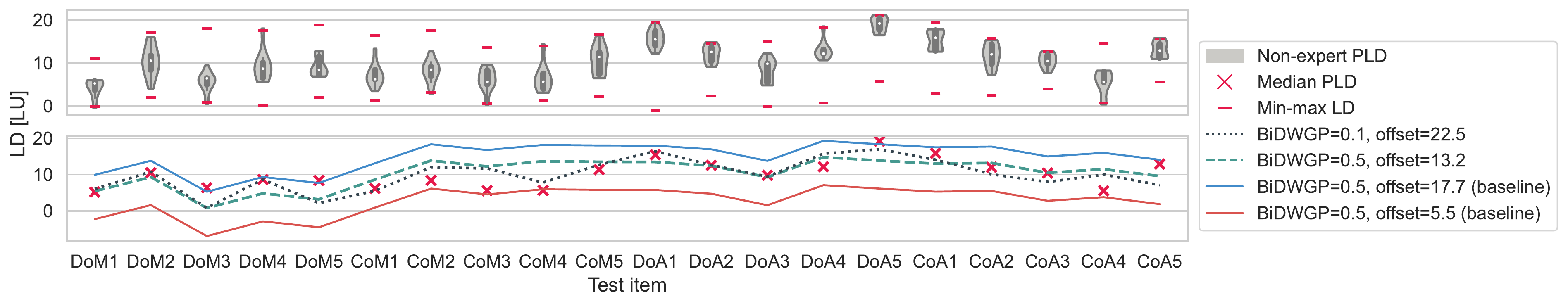}
\vspace{-20pt}
\caption{Top: Violin plot of non-expert PLDs. The red dashes represent the item-dependent maximum and minimum LDs available during the listening test as measured after BSS Eval signal decomposition. Bottom: BiDWGP-based predictions of median PLDs using baseline and proposed parameters.}
\vspace{-12pt}
\label{fig:predPlot} 
\end{figure*}

\subsection{Listening Test Results}
\label{section:testResults}

An Analysis of Variance (ANOVA) was performed on the PLDs obtained in the listening test (Table~\ref{table:anovaTable}). A heteroscedasticity-consistent covariance matrix \cite{long2000using} was employed, and $4$ factors were considered in this analysis: \emph{participant}, \emph{item}, \emph{foreground-background type} (CoA, CoM, DoA, DoM), and \emph{experience}. Of the statistically significant factors ($\alpha = 0.05$), \emph{participant} explains the largest share of the total variability, at $32.31\%$, with an average interquartile range across subjects of $5.7\mskip3mu$LU. This implies that due to the considerable influence of individual listening preferences, a ``one-size-fits-all'' mixture may not be attainable. Even so, median PLDs may be predicted for a given program to offer a preset that can suit most listeners, while others would still have the option of personalizing the LD. Next, \emph{item} and \emph{foreground-background type} displayed the second and third largest effect sizes, explaining $27.18\%$ and $24.15\%$ of the total variability, respectively. This suggests that PLDs could be predicted in an item-dependent fashion. Specifically, the large effect size for \emph{foreground-background type} indicates that classifying items by their combination of FG (commentary or on-set dialogue) and BG (music or ambience) could be a promising approach.

Figure~\ref{fig:pldByBgAndExp} reports recorded PLDs grouped by item class to highlight the class-dependent nature of PLDs. Both expert and non-expert subjects favored greater BG attenuation for items featuring ambience rather than music. Tukey's HSD post-hoc test indicated that on average, PLDs are approximately $5\mskip3mu$LU higher for ambience than for music ($p<0.001$). This difference is consistent with what has been observed in \cite{torcoli2019preferred}.  Our work introduced on-set dialogue as a new class of item which had not been considered in previous works about PLDs. On average, non-experts favored LDs $2.3\mskip3mu$LU higher for DoA than for CoA ($p<0.001$), while no significant difference was found between DoM and CoM ($p=0.85$). Specific reasons for this observation are beyond the scope of this work.

\begin{table}[t!]
\centering
\renewcommand{\arraystretch}{1.1}
\vspace{1pt}
\caption{Target BiDWGP, offset and MAE for PLD predictions using baseline and proposed parameters (in bold).}
\begin{tabular}{c c c}
\toprule
\textbf{BiDWGP} & \textbf{Offset} & \textbf{MAE}\\
\hline
$0.5$ & $\phantom{0}5.5\mskip3mu$LU &	$7.5\mskip3mu$LU \\\relax
$0.5$ & $17.7\mskip3mu$LU &	$5.0\mskip3mu$LU \\\relax
$\mathbf{0.5}$ & $\mathbf{13.2}\mskip3mu$\textbf{LU} &	$\mathbf{3.2}\mskip3mu$\textbf{LU} \\\relax
$\mathbf{0.1}$ & $\mathbf{22.5}\mskip3mu$\textbf{LU} &	$\mathbf{2.5}\mskip3mu$\textbf{LU} \\\bottomrule\relax 
\end{tabular}
\vspace{-22pt}
\label{table:estFitParams}
\end{table}

\section{Predicting Preferred Loudness Differences}

Two previous works using the BiDWGP OIM to predict PLDs were used as baselines for the present study and are hereby referred to as $0.5+5.5\mskip3mu$LU \cite{tang2018automatic} and $0.5+17.7\mskip3mu$LU \cite{torcoli2019preferred} respectively. Both sets of parameters were tested with the data gathered for our experiment, as reported in Table \ref{table:estFitParams}. $0.5+5.5\mskip3mu$LU was found to underestimate median PLDs in most cases with a Mean Absolute Error (MAE) of $7.5\mskip3mu$LU, while $0.5+17.7\mskip3mu$LU resulted in an overestimation for $80\%$ of the dataset (MAE = $5\mskip3mu$LU).

Two new sets of improved prediction parameters were computed, which are also depicted in Table \ref{table:estFitParams}.
Non-expert median PLDs were predicted using the BiDWGP OIM. The prediction was performed on non-expert data, as they are more representative of general audiences who would most benefit from DE. The BiDWGP OIM outputs a real value $\in [0,1]$, with $0$ corresponding to completely unintelligible speech and $1$ to full intelligibility. A mapping between BiDWGP and PLDs was established by first computing BiDWGP scores for target $\xi$, starting at $0$ up to $1$ in increments of $0.1$. For each of the test items, the BG is attenuated until the corresponding BiDWGP output falls within $\pm\mskip3mu0.01$ of target $\xi$. An offset $\epsilon$ was then added to the predicted LD. The predicted PLD is then selected as the LD corresponding to the target and offset pair that minimizes the MAE with respect to the median non-expert PLDs recorded in the listening test. More formally, the optimum BiDWGP score $\xi^{*}$ and offset $\epsilon^{*}$ corresponding to the predicted PLD $\widehat{L}$ are computed across each of the $Q$ items as:
\vspace{-8pt}
\begin{equation}
\xi^{*}, \epsilon^{*} = \argmin_{\xi, \epsilon }  \frac{\mathrm{1} }{\mathrm{Q}}  \sum_{q=1}^{Q}  \left\lvert (\widehat{L}(\xi, \epsilon) - L_{\textrm{ground-truth}}) \right\rvert. 
\end{equation}
\vspace{-8pt}

Non-expert median PLDs and their prediction are shown in Fig.~\ref{fig:predPlot}, along with the range of actual LDs available during the listening test. First, we fixed $\xi=0.5$ to match the baselines, and found that $\epsilon=13.2\mskip3mu$LU minimizes the MAE. When $\xi$ is allowed to vary, the optimum parameters are $\xi=0.1$ and $\epsilon=22.5\mskip3mu$LU. While the ANOVA suggests that PLDs are class-dependent we opted not to perform class-specific prediction with our data, in order to avoid overfitting. Future works should test this approach with a larger dataset. Both proposed parameter sets display a reduced MAE with respect to the baseline methods. Furthermore, it is worth noting that the proposed parameters yield predictions that lie between the two baseline predictions for all items. As such, the baselines appear to provide upper and lower bounds for prediction performed on dialogue-separated signals.

\section{Conclusion}

This work considered Preferred Loudness Differences (PLDs) between dialogue and background sounds for the application of remixing dialogue-separated signals. A listening test considering an expanded range of broadcast audio items provided insights into what factors influence PLDs. Most of the variability in the data was explained by individual preferences, confirming the need for audio personalization. Average PLDs are $2.3\mskip3mu$LU higher for on-set dialogue than for commentary when the background features ambience. PLDs are also highly dependent on individual items. Our work showed that prediction of median PLDs using Objective Intelligibility Metrics (OIMs) can be carried out even in the presence of dialogue-separated signals. A set of optimum prediction parameters was proposed, resulting in mean absolute errors at least $36\%$ lower than existing baselines. The two baselines provide possible lower and upper bounds for prediction.
Future works should validate our results, quantify the effect of varying degrees of distortions introduced by dialogue separation, and explore the potential for class-specific prediction on a larger dataset. Furthermore, given the high computational cost of the BiDWGP metric, efficient implementations are needed to employ OIM-based prediction in practice.

\balance
\bibliographystyle{./IEEEtran}
\bibliography{./IEEEexample}

\end{document}